# Tunable Quantum Tunneling of Magnetic Domain Walls

J. Brooke*[†], T.F. Rosenbaum* & G. Aeppli[†]

*The James Franck Institute and Department of Physics, The University of Chicago, Chicago, IL 60637

[†]NEC Research Institute, 4 Independence Way, Princeton, NJ 08540

Perhaps the most anticipated, yet experimentally elusive, macroscopic quantum phenomenon[1] has been spin tunneling in a ferromagnet[2], which may be formulated in terms of domain wall tunneling[3,4]. One approach is to focus on mesoscopic systems where the number of domain walls is finite and the motion of a single wall has measurable consequences. Research of this type includes magnetotransport measurements on thin ferromagnetic wires[5] and magnetization experiments on single particles[6,7], nanomagnet ensembles[8-10], and rare earth multilayers[11]. A second method is to investigate macroscopic disordered ferromagnets[12-15], whose dynamics are dominated by domain wall motion, and search the associated relaxation time distribution functions for quantum effects. Both approaches have revealed clear deviations from thermal relaxation in the form of finite timescales that persist as temperature T approaches zero. But while the classical, thermal processes in these experiments are easily regulated via T, the quantum processes have not been tunable, making definitive interpretation in terms of tunneling difficult. Here we report on a disordered magnetic system for which it is possible to adjust the quantum tunneling probabilities with a knob in the laboratory. We are able to model both the classical, thermally activated response at high **temperatures and the athermal, tunneling behavior at low temperatures within a simple, unified framework.**

Fig. 1a depicts domain wall motion in the classical and quantum limits against the background of a fixed potential landscape, which pins the walls. In the classical case, highlighted in blue, the domain wall moves *over* the potential barrier, thermally flipping spins as it advances. In the extreme quantum case (red arrow), the domain wall tunnels *through* the barrier, with the possibility in rare instances of flipping all barrier spins simultaneously. The problem illustrated in Fig. 1a reduces to the more familiar quantum barrier tunneling problem in Fig. 1b if the domain wall is associated with a particle of effective mass m, moving in a one-dimensional potential derived from the real three-dimensional pinning potential. The mass is a key parameter – heavy particles behave classically and remain more localized than light particles. Our experiments on the disordered ferromagnet $LiHo_{0.44}Y_{0.56}F_4$ demonstrate that Fig. 1b provides a detailed description of domain wall motion, with m a continuously tunable parameter.

$LiHoF_4$ is a tetragonal insulating ferromagnet with a Curie temperature $T_c = 1.53$ K and an ordered moment along the crystal c-axis. For magnetic fields $H_t$ applied perpendicular to the c-axis, the material becomes the experimental realization of the simplest quantum spin model, the Ising ferromagnet in a transverse field. The corresponding Hamiltonian is

$$H = -\sum_{i,j}^{N} J_{ij} \boldsymbol{S}_i^z \boldsymbol{S}_j^z - \Gamma_e \sum_{i}^{N} \boldsymbol{S}_i^x \ , \qquad (1)$$

where the $\sigma$'s are Pauli spin matrices located at lattice sites i and j, the $J_{ij}$'s are longitudinal couplings, and $\Gamma_e$ is an effective transverse field, perpendicular to the Ising axis and proportional to $H_t^2$ for small $H_t$. In the $\Gamma_e = 0$ limit, atomically thin domain walls separate regions with $\sigma_z = +1$ from $\sigma_z = -1$. This wall is dynamically stable for $\Gamma_e = 0$ because H commutes with $\sigma_z$; the domain wall has infinite mass. For non-zero $\Gamma_e$, the commutator [H, $\sigma_z$] no longer vanishes and wall motion can occur; the domain wall mass is now finite. Eventually, when $\Gamma_e$ becomes comparable to J, there is a quantum critical point (occurring at $\Gamma_c$ = 5.3 K for pure LiHoF$_4$)[16] beyond which the Ising ferromagnetism is unstable. As $\Gamma_e \rightarrow \Gamma_c$, the domain walls broaden to fill the entire system (resulting in zero net moment), or equivalently, their mass approaches zero. In practice, even for zero applied $\Gamma$, the internal dipole fields of LiHoF$_4$ create an internal field $\Gamma_i$ which slightly broadens the domain walls[17] and cuts off the mass at large but finite value. The effective magnetic field is thus $\Gamma_e = \Gamma + \Gamma_i$. If suitable barriers to domain wall motion are introduced into LiHoF$_4$, the domain wall dynamics can pass from classical (large m) to quantum mechanical (small m) limits within our measurement window via a simple increase of $\Gamma$ from 0 towards $\Gamma_c$. We can insert such barriers by the random, partial substitution of non-magnetic Y for magnetic Ho, leading to quenched disorder ideal for pinning domain walls. With suitable magnetic dilution x, LiHo$_x$Y$_{1-x}$F$_4$ in a transverse field is a macroscopic system for which we can vary domain wall mass for a potential energy landscape dominated by fixed pinning centers.

We employ both static and dynamic measurements to explore the nature of the ordered state. dc magnetometry reveals standard magnetization – longitudinal field (M-H) hysteresis loops characteristic of pinned domains in a ferromagnet, with widths of order tens of Oersteds and a saturation magnetization $4\pi M \sim 200$ Oe (Fig. 2). Both raising temperature (thermal fluctuations) and increasing the transverse field (quantum fluctuations) serve to depin the domain walls and narrow the hysteresis loops.

The technique we use to probe domain wall motion is magnetic susceptometry, which measures the incremental changes in magnetization due to an infinitesimal oscillating field. For a ferromagnet, such changes correspond to the growth of domains polarized parallel to the field at the expense of domains with antiparallel polarization. The growth occurs via domain wall motion of the type illustrated in Fig. 1a. If the motion is fast on the scale of the measuring frequency $f$, the susceptibility $c(f)$ will be frequency-independent, while, if it is slow, the domains will not be able to follow the ac field and the in-phase part of $\chi$ will be reduced. The simplest motion is relaxational, and is characterized by a single relaxation time $\tau$. For classical barrier hopping, $\tau$ will follow a thermally activated Arrhenius form, $1/t = f_m \exp(-\Delta/T)$, with microscopic attempt frequency $f_m$ and barrier energy $\Delta$. In the quantum limit, the rate at which the wall can tunnel through the same barrier will be the squared amplitude of the wavefunction on the other side of the barrier a distance $w_o$ away. For a square barrier, this is

$$1/t = f_m e^{-2w_o \sqrt{\frac{2m}{h^2}\Delta}} \ .$$

(2)

Thus, by measuring $c(f)$ as a function of T, one can monitor the evolution from classical Arrhenius behavior to the T-independent quantum regime. For the random ferromagnet $LiHo_{0.44}Y_{0.56}F_4$, the domain wall dynamics will be defined not by a single $\tau$, but by a distribution $\rho(\tau)$ of relaxation times, associated with a distribution of barriers $\Delta$, as illustrated in Fig. 1c. Assuming that the different $\tau$'s are due to independent (non-interfering) processes, the response function becomes

$$c(f) = \int_0^\infty r(t) c_t(f) dt \,,$$

(3)

where $c_t(f) = c_\circ / (1 - 2p\acute{i}ft)$ is the Debye response[18]. Our strategy, therefore, is to measure $c(f)$, and then to examine the implications for the evolution of $\rho(\tau)$ with T and $\Gamma$.

We plot in Fig. 3a the phase diagram as a function of temperature and transverse field for single crystal $LiHo_{0.44}Y_{0.56}F_4$. In the large T, small $\Gamma$ classical limit, the system is a disordered ferromagnet with a Curie temperature, $T_c(\Gamma=0) = 0.65$ K, while for T=0 it has a quantum critical point at $\Gamma_c(T=0) = 1.6$ K[19]. The domain wall dynamics are encoded in the frequency-dependent magnetic susceptibility data of Fig. 3b. To analyze our results, we use Eq. (3) and choose $\rho(\tau)$ to be as simple as possible yet consistent with the observed data:

$$r(t) = r_o \, d(t) + r_1 t^{-1} \Theta(t - t_l) \Theta(t_h - t) \ . \tag{4}$$

In Eq. (4), prefactors $\rho_o$ and $\rho_1$ provide normalization, while the $1/\tau$ form of the second term yields the observed logarithmic divergence, with cutoffs at the low and high ends, $\tau_l$ and $\tau_h$, dictated by the data ($\tau_h$ is only observable for a few curves at the highest T and $\Gamma$). The delta-function at very short times accommodates the flat spectral response at high frequencies.

Given the simplicity of $\rho(\tau)$, the parameters $\tau_l$ and $\tau_h$ summarize the entire measured dynamics of LiHo$_{0.44}$Y$_{0.56}$F$_4$, and consistently fitting the data to this form yields unique values for f$_o$ whose systematic behavior as a function of $\Gamma$ and T can then be studied. We focus on the crossover frequency $f_o = 1/\tau_h$, corresponding to the fastest large-scale relaxation, and the most weakly-pinned domain walls. Plotting $f_o$ against inverse T for several transverse fields yields Fig. 4, the central result of our experiment, which demonstrates explicitly the evolution from classical to quantum relaxation. At the higher temperatures, the domain wall relaxation follows the Arrhenius form with a universal ($\Gamma$- and T-independent) microscopic attempt frequency, $f_m = 2.2 \pm 0.2 \times 10^5$ Hz, and a $\Gamma$−dependent barrier energy $\Delta$, shown in Fig. 1e. For increasing transverse field, the system approaches the disordered paramagnetic state, with the result that $\Delta$ is reduced linearly, $\Delta(\Gamma) = \Delta_0 (1 - \Gamma / \Gamma_\Delta)$ with $\Delta_o = 0.54 \pm 0.03$ K. The solid line through the data corresponds to this form, and extrapolates to zero at $\Gamma_\Delta = 1.3 \pm 0.2$ K, remarkably close to the quantum critical field $\Gamma_c(T=0)$ beyond which the material is paramagnetic (Fig. 3a). Below T ∼ 0.1 K, there are clear deviations from Arrhenius behavior, with $f_o$ becoming temperature-independent as T → 0. The crossover temperature to quantum behavior increases with $\Gamma$; at high transverse field the increased tunneling probability permits the quantum regime to extend to higher T. This observation, combined with the continued evolution down to the lowest temperatures of the $\Gamma$ – T phase diagram of Fig. 3a, proves that the saturation in $f_o$ is intrinsic and not due to a thermal decoupling of the sample from the dilution refrigerator.

The transverse field is far more efficient at relaxing the system in the quantum regime than in the thermal regime. The comparative advantage is most apparent in Fig.1d, where we show $f_o$ as a function of $\Gamma$ at several T. At the T = 0.030 K base temperature, tripling $\Gamma$ from 0.19 K to 0.58 K increases $f_o$ by one and a half orders of magnitude, while at T = 0.150 K it has a negligible effect. Having determined the barrier $\Delta$ at each $\Gamma$ from the high temperature regime, we are able to use the barrier tunneling form Eq. (2) to extract the combination $mw_o^2$ from the measured $f_o$. It is clear that $mw_o^2$ varies more rapidly with $\Gamma$ than does $\Delta$ (Fig. 1e); the principal effect of the transverse field in this part of the $\Gamma$-T plane is to reduce the effective mass of the domain walls. The barrier width $w_o$, which characterizes the distances over which the domain walls must hop, should be fixed by the frozen impurity configuration and ought to be independent of $\Gamma$. Hence, we attribute the reduction of $mw_o^2$ with $\Gamma$ to a reduction in m, describable by $m = \lambda /(\Gamma + \Gamma_i)$, where $\lambda$ is a constant (solid line in Fig. 1e).

The simplest expression for the entire T- and $\Gamma$-dependent relaxation is based on assuming that classical and quantum processes provide independent relaxation channels. Thus, we simply add the quantum and classical forms identified above, i.e.

$$f = f_m \left\{ \exp\left(- \Delta_o \left(1 - \Gamma/\Gamma_\Delta\right)/T\right) + \exp\left(- A\sqrt{\left(\Gamma + \Gamma_i\right)^{-1}} \sqrt{\Delta_o \left(1 - \Gamma/\Gamma_\Delta\right)}\right)\right\}.$$

(5)

Here, $f_m$ characterizes the response time of the most weakly-pinned domain walls and $A = 2w_o \sqrt{2\lambda / \hbar^2}$. This physically transparent expression, while only first order in the most relevant parameters of the system and lacking cross terms, fits the data quite well in both T and Γ as evidenced from the solid lines of Fig. 4. Naturally, it is only valid deep in the ordered phase of Fig. 3a, with open circles denoting where measurements were taken.

A further test of the tunneling and hopping particle model for domain wall motion (illustrated in Fig. 1b and encapsulated by Eq. (5)) is whether it applies to more than just the fastest processes, derived from the lowest barriers. In particular, can a single mass account for the entire spectral response in both the classical and quantum regimes exhibited for a single Γ in Fig. 3b? Indeed it can, as evidenced by the solid lines through the data. This three-parameter fit is derived from a fixed barrier distribution function $\rho(\Delta)$, shown in Fig. 1c, such that for each $\Delta$ and T, the relaxation time is given by Eq. (5). For the smallest $\Gamma = 0.19$ K, the distribution is narrow, with a width of the same order as the centroid. This sharpness at low Γ is consistent with ferromagnetic ordering, in spite of the significant (56%) disorder. Increasing Γ broadens the barrier distribution.

We have discovered a ferromagnet with tunable quantum tunneling of domain walls. A very simple model describing the domain wall as a particle with fixed mass moving via quantum tunneling or thermal hopping among minima in a random potential provides an excellent description of data collected over four decades of frequency. In contrast to tunneling involving bare particles such as electrons or protons, the mass is a continuous variable, adjusted simply by an external field. One very important question concerns the number N of spins which are tunneling coherently – have we seen macroscopic quantum tunneling? The collective nature of the spin dynamics is borne out by the numbers. An individual (isolated or weakly coupled) spin would have a flipping rate of order $\Gamma/h$, which for $\Gamma = 0.4$ K is $10^{10}$ Hz, five orders of magnitude larger than the measured $f_o$. More directly, we can estimate N from the measured mass (Fig. 1e) of the domain walls. In particular, if we consider a one-dimensional Ising model subject to a net transverse field $\Gamma_e$ much less than the ferromagnetic coupling, the domain wall is a particle of mass m=$\hbar^2/2a^2\Gamma_e$ where $a$ is the lattice spacing. (This may be obtained from the semi-classical form $d^2E/dk^2 = \hbar^2/m$ applied to the domain wall dispersion of the transverse Ising model[20], and is the quantum spin-1/2 limit of the Döring mass[21,22].) For an array of N parallel chains the mass will be N times larger, i.e. m = N$\hbar^2/2a^2\Gamma_e$. Using the measured masses in Fig. 1e, we estimate that wall segments containing N≈10 spins tunnel together, and conclude that quantum relaxation in LiHo$_{0.44}$Y$_{0.56}$F$_4$ is coherent on the nanometer scale.

## Methods

We suspended a needle-shaped cylinder of LiHo$_{0.44}$Y$_{0.56}$F$_4$ (with aspect ratio 8 to minimize demagnetization effects) from the mixing chamber of a helium dilution refrigerator into the bore of an 8 T superconducting magnet oriented perpendicular to the crystalline c-axis (to within 0.5$^\circ$). A trim coil along the Ising direction nulled any unwanted longitudinal field component. The ordered state was always entered by cooling in large transverse field $\Gamma$ and zero longitudinal field to the target temperature and then reducing $\Gamma$ through the phase boundary. Static measurements (i.e., data of Fig. 2) were obtained using 200μm × 80μm thin film InAs Hall probes, crafted for low temperature use[23], placed perpendicular to the Ising axis on the end of the sample cylinder. The dynamic response was measured after 12 hours of equilibration time through the complex ac susceptibility, $c(f) = c'(f) + ic''(f)$, along the Ising axis with a standard gradiometer configuration, using digital lock-in amplifiers for the reference and signal channels. The energy splitting $\Gamma$ between the originally degenerate Ising doublet (Eq. (1)) is calculated from the laboratory transverse magnetic field $H_t$ using the known crystal field levels of the Ho$^{3+}$ ion[24].

We are grateful to D. Bitko, S. Girvin, S. Nagel, P. Stamp, and T. Witten for enlightening discussions.  The work at the University of Chicago was supported primarily by the MRSEC Program of the National Science Foundation.

**Correspondence should be addressed to T.F.R.**

**e-mail: t-rosenbaum@uchicago.edu.**

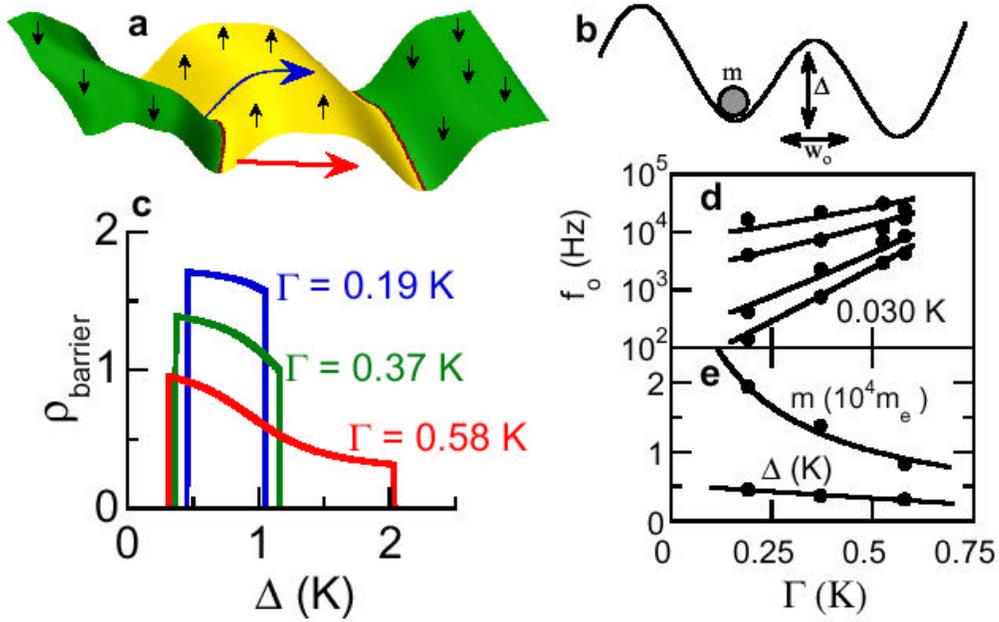

Figure 1:  Domain wall tunneling.  **a**, Cartoon depicting motion of a domain wall separating regions of opposing spin orientation.  The classical, thermally-activated process is indicated with a blue arrow, and the quantum tunneling route is shown in red.  **b**, Sketch of the domain wall in **a** modeled as a particle in a one-dimensional potential with barrier height $\Delta$ and width $w_o$.  **c**, Experimentally obtained domain wall barrier distribution as a function of energy at a series of transverse fields, $\Gamma$.  **d**, Frequency response of the weakest-pinned domain walls as a function of $\Gamma$ at temperatures T = 0.030, 0.070, 0.110, and 0.150 K, in order of increasing response magnitude.  Solid lines follow from Eq. (5).  **e**, Best-fit values to Eq. (5) for the domain wall mass m and potential barrier height $\Delta$ of Eq. (2) for the weakest-pinned domain walls, as functions of $\Gamma$.  The solid line through m is $\lambda/(\Gamma + \Gamma_i)$, where $\lambda = 0.66 \pm 0.05 \times 10^4$ $m_e$ K and $\Gamma_i$ = 0.15±0.02 K.  To obtain real masses, the tunneling distance, $w_o$, has been set to the average spacing *a* between magnetic ions, 8.1 Å.

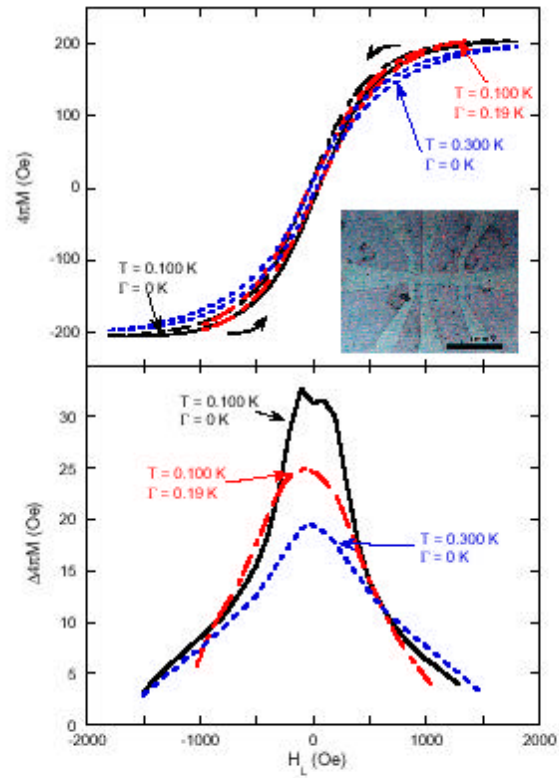

Figure 2: Static magnetization. **a**, Magnetization – Longitudinal Field (M-H) hysteresis loops after zero longitudinal field cooling at three temperatures and transverse fields, consistent with domain ordering: T = 0.300 K, Γ = 0 K (blue short dash); T = 0.100 K, Γ = 0 K (black solid line); T = 0.100 K, Γ = 0.19 K (H$_t$ = 5 kOe) (red long dash). **inset,** Photograph of the InAs Hall bar assembly used for the magnetization measurements. **b**, The M-H loop widths of the curves in panel **a**.

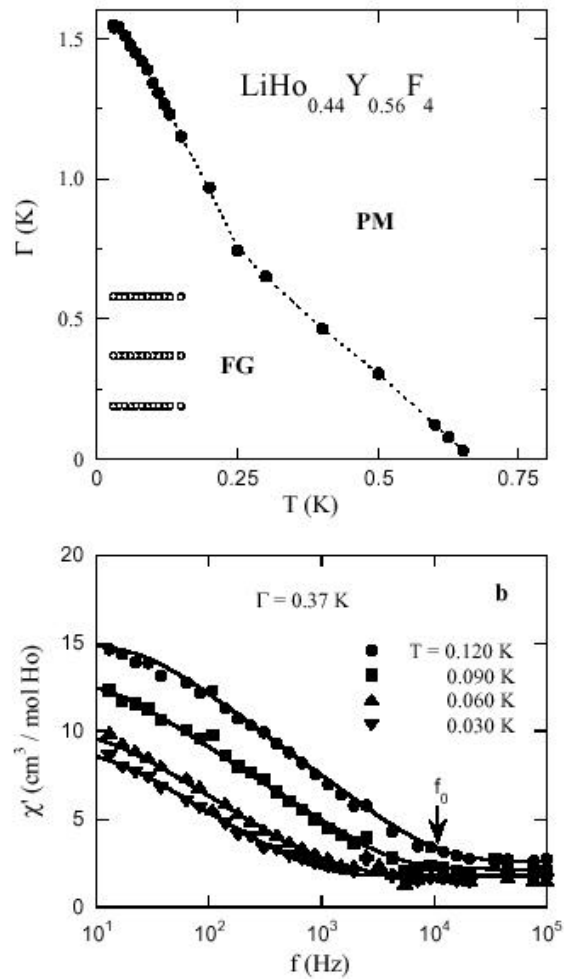

Figure 3: Phase diagram and spectral characterization of the ordered state. **a**, Transverse field Γ – Temperature T phase diagram for the disordered Ising magnet, with 44% of the sites occupied by (holmium) magnetic dipoles. The transverse field introduces tunneling modes for magnetization that depress the temperature for spin freezing in a controllable fashion. Filled circles denote the phase boundary, measured by the cusp in susceptibility at 5 Hz, with the dashed line a guide to the eye. Open circles denote the values of Γ and T at which the susceptibility was measured for investigating domain wall tunneling. PM = Paramagnet, FG = Ferroglass. **b**, The spectroscopic response in the quantum limit is characterized by a logarithmic dependence of the real part of the magnetic susceptibility, χ', over several decades below a characteristic

frequency $f_o$. An arrow highlights this crossover for T=0.120 K in the figure. Reducing T decreases $f_o$, as well as the amplitudes of both the logarithmic low-$f$ and constant high-$f$ terms. However, the effect of reducing T (i.e. by similar increments $\delta$T) diminishes with decreasing T. Thus, the magnetic relaxation appears to approach a T-independent quantum limit on all measured frequency scales. Points are from measured data at $\Gamma$ = 0.58 K, with error bars smaller than the symbol sizes, and lines are best-fit values to Eqs. (3) and (4).

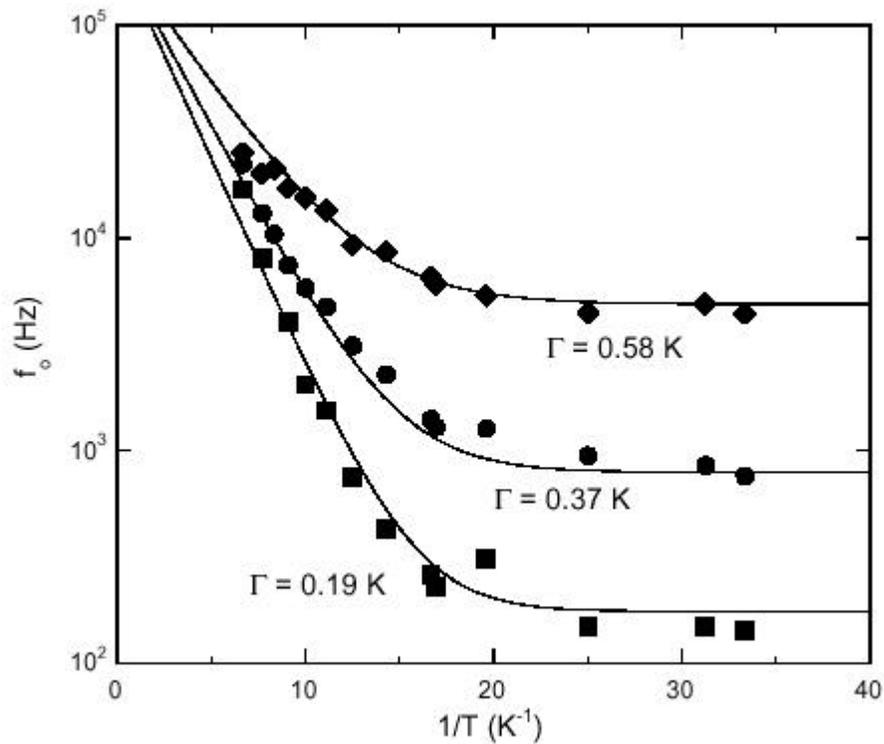

Figure 4: The characteristic frequency for magnetic domain relaxation as a function of both classical (T) and quantum (Γ) variables (data) along with the best fit of Eq. (5). At high temperature T, the relaxation is thermally activated over energy barriers that decrease with increasing transverse field Γ. Below T ~ 0.1 K, the system smoothly enters a temperature-independent tunneling regime of simple barrier tunneling character. Error bars are comparable to the symbol sizes.